\\

Title: Radionuclide transport in the Yenisei River

Authors: S.M. Vakulovsky, E.G. Tertyshnik and A.I. Kabanov

Data characterizing the pollution of the Yenisei River (water and bottom sediment) by radionuclide resulting from the use of the river's water for cooling industrial reactors in the Mining-Chemical Complex are presented. Studies have been made of the contamination of the river during the period when reactors with direct flow cooling were used and after these were shut down. Distinctive features of the migration of radionuclide in the Yenisei are noted, in particular, their distribution between the solid and liquid phases. The amounts of $^{137}$Cs, $^{65}$Zn, $^{60}$Co, $^{54}$Mn, and $^{152}$Eu in the channel are determined from the effluent discharge site to Dudinka. The rate of continuous self removal of $^{137}$Cs is estimated to be µ=0.19 1/yr (corresponding to a half purification time of 3.6 years) for a 600-km long segment of the river bed.

\\

**Radionuclide transport in the Yenisei River**

S.M. Vakulovskii, E.G. Tertyshnik and A.I. Kabanov

This paper describes the migration of radionuclidesin the Yenisei River from the place where they discharged as effluent from the Mining-Chemical Complex near Krasnoyarsk to Dudinka. Since 1958, this complex, located on the right bank of the Yenisei 50 km from Krasnoyarsk, has used the river's waters for cooling of industrial nuclear reactors for the production of $^{239}$Pu. River water was passed through the cooling system of the reactors and returned to the Yenisei. Significant amounts of radionuclides were present in the effluent; these were mainly formed by neutron activation of impurities contained in the river water. In 1992, the direct-flow reactors were shut down. As a result, the activity of the river water has decreased by hundreds of times.

Since 1972, expeditions have been mounted for studying the radioactive contamination of the water, bottom segments, and shore floodplains of the Yenisei owing to the activities of the Complex. These studies were continued in 2000 and 2001, after the direct-flow reactors had been shut down. Studies of the distribution of radionuclide in the Colombia River [1], whose water was used for cooling the industrial reactors at Hanford (USA), were of some use for planning the experiments, choosing the techniques for concentrating the radionuclide, and calculating the activity of the water in the channel of the Yenisei.

Samples of water, suspensions, bottom sediments, and shore soils were taken in a segment of the river of approximately 2000 km in length from Krasnoyarsk to Dudinka. At fixed cross sections, located at distances of 268, 419, and 876 km from Krasnoyarsk (Table 1), samples were taken at various rates and indifferent seasons of the year and suspensions were separated by means of multisection filtering devices [2, 3].

Table 1. Location of Cross Sections and Sampling Points in the Yenisei River

| Section № | Distance from Krasnoyarsk, km | Closest population center |
|---|---|---|
| - | 86 | Atamanovo (waste discharge point) |
| 1 | 266 | Kazachinskoe |
| 3 | 327 | Shirokii Log |
| - | 330 | Strelka (mouth of the Angara River) |
| 5 | 414 | Yeniseisk |
| 2 | 864 | Bor (southerly mounth of the Podkamennaya Tunguska River) |
| - | 1446 | Turukhansk |
| - | 1745 | Igarka |
| - | 2000 | Dudinka |

The amounts of γ–emitting radionuclide in the samples were determined in a stationary laboratory using gamma spectrometers equipped with Ge(Li) detectors of Russian manufacture and LP-4840 and LP-4900 multichannel analyzers from Nokia (Finland). The methods for concentrating the radionuclide from the water and for the γ-analysis are described elsewhere [4]. The amount of short lived $^{24}$Na in the water was determined under field conditions using a scintillation spectrometer with a 63×63 mm NaI (Tl) crystal. Because of the comparatively low amount of radionuclide in the Yenisei water, the sample volumes of river water had to several hundred liters in order to obtain reliable results. Suspensions were separated from the water using filtration devices. Dense paper filters ("blue ribbon" type) were used as the filter material and prefilters made of FPP-15-1,5 filter tissue were installed above them. The filters, along with the suspensions separated from them, were dried and then incinerated. After weighing, the ash was subjected to radioisotope analysis. Water samples with volumes of 100 liters were evaporated to determine the total activity in solutionand (liquid) and the solid phase. As model experiments in the laboratory showed, there was almost no loss of radionuclide during evaporation when the mineral content of the sample water was increased artificially, stable carriers of the radionuclide were introduced, humic acids were added, and a suitable evaporation regime was instituted [5].

The data in Table 2, averaged over 10 and 7 samples with water flows of 2930 and 7600 m$^3$/sec during the measurements for sections 1 and 2, respectively, indicate that most of the radionuclide are transported predominantly in the solid phase in the Yenisei waters (f > 0.5), while $^{51}$Cr and $^{24}$Na migrate primarily in dissolved form (f < 0.5), and $^{137}$Cs is distributed equally between the solid and liquid phases (f = 0.5).

Table 2. Average volume activity of radionuclide in water $A_v$, measured in the Summer of 1979 and average fraction of radionuclide in the suspension f

| Radionuclide | $T_{1/2}$, days | $A_v$ Bq/m$^3$ | δ, % | f | δ$^*$ | $A_v$ Bq/m$^3$ | δ, % | f | δ |
|---|---|---|---|---|---|---|---|---|---|
| | | Section 1 | | | | Section 2 | | | |
| $^{152}$Eu | 4858 | 6 | 0.20 | 0.6 | 0.21 | 2.2 | 25 | 0.8 | 30 |
| $^{51}$Cr | 27.7 | 6060 | 0.04 | 0.12 | 0.1 | 2040 | 10 | 0.1 | 10 |
| $^{137}$Cs | 10958 | 8.7 | 0.15 | 0.5 | 0.17 | 4.6 | 12 | 0.5 | 13 |
| $^{58}$Co | 70.9 | 49 | 0.07 | 0.85 | 0.06 | 14 | 8 | 0.6 | 12 |
| $^{54}$Mn | 312.2 | 25 | 0.04 | 0.9 | 0.14 | 8.6 | 12 | 0.7 | 14 |
| $^{46}$Sc | 83.8 | 26 | 0.09 | 0.72 | 0.09 | 7.1 | 14 | 0.6 | 13 |
| $^{65}$Zn | 244 | 46 | 0.06 | 0.53 | 0.12 | 11 | 15 | 0.7 | 20 |
| $^{59}$Fe | 44.5 | 41 | 0.12 | 0.9 | 0.15 | 20 | 10 | 0.4 | 20 |
| $^{60}$Co | 1925 | 10 | 0.05 | 0.93 | 0.1 | 4.1 | 9 | 0.6 | 15 |
| $^{32}$P | 14.3 | 1640 | 0.12 | 0.4 | 0.5 | 300 | 20 | 0.7 | 50 |
| $^{24}$Na | 0.61 | 19000 | 0.06 | - | - | | | | |
| δ$^*$ is the relative error | | | | | | | | | |

Sometimes, the behavior of the radionuclide in the two phase water-suspension system is characterized using the distribution coefficient for the suspension, $K_d$ is used. Evidently, the fraction of the radionuclide in the suspension is related to the distribution coefficient by

$$K_d = 10^3 f/(1-f)M,$$

where M is the amount of suspended matter in the water (turbidity), g/liter. In this paper, the dimensionless coefficient f is used, since:

- the amount of suspended matter in the Yenisei water is extremely small and the increase in the mass of a filterowing to deposited suspension was often not determined reliably;
- the fraction of radionuclide fixed in the largest fractions of the suspension is small and the mass of these fractions is a significant part of the overall mass of the suspensions, so that the distribution coefficients are subject to fluctuations over relatively larger limits than is the coefficient f as the hydrological state of the river varies. Thus, the "damming" effect noted in Ref. 6 is a consequence of the setting of large suspensions on the bottom, so that in sections of the river with low flows the distribution coefficient increases, while the slowing down in the velocity does not affect the fraction of a radionuclide in the suspension;
- sorption of $^{137}Cs$ by a suspension and bottom sediments is irreversible, while use of the distribution coefficient assumes the existence of a concentration equilibrium in the water-suspension or water-bottom sediment system.

The distribution of several radionuclide between the solid and liquid phases in Yenisei water have been expressed in terms of the distribution coefficient [7]. These data vary over wide limits and it is impossible to establish whether they depend on distance. It is also impossible to compare the results on the interphase distribution of radionuclide given here with the data reported there [7]. At the same time, some general behavior of the radionuclide migration in the Columbia [1] and Yenisei Rivers has been established by characterizing the distribution of radionuclide in the solution-suspension system in terms of the fraction of radionuclide in suspension.

The migration of stable micro- and macroelements in a river flow is characterized by a certain relationship between the physical and chemical forms of each element, including the fractions determined to be associated with a suspension. Using river water to cool nuclear reactors disrupts this equilibrium because of the water preparation processes and because of chemical reactions accompanying the nuclear reactions by which the radionuclide are created. Measurements from 1975 and the data of Ref. 1 indicate that radionuclides are present in the discharge water mainly in dissolved form (low values of f). After the effluent enters the river, the form of the radionuclide undergoes a transformation; ultimately a substantial fractions of them end up associated with a suspension. Probably the radionuclide are converted to states that are typical for the corresponding stable elements in a given river. Near cross section 1, which is 188 km away from the point where the effluent is discharged into the river, the transformation of the physical and chemical states of the radionuclide seems to have come to completion, since the values of f obtained at this section are the same as at sections 3 and 5 (Table 1). The fact that the fraction of radionuclide in the suspension is constant for a steady discharge regime and is independent of the hydrological state makes it possible to monitor the radioactive contamination of the river just by collecting samples of the suspensions with the aid of a high capacity filtration system and to calculate the total activity using the corresponding fractions of the radionuclide in the suspension.

At section 2, which is 796 km from the discharge point, the fraction of radionuclide in the suspension decreases by roughly 20% compared to the characteristic values at sections 1, 3, and 5 for most of the radionuclide; the statistical significance of this observation was established from 10 samples (see Tables 2 and 3). The stable fraction in the suspension at sections 1, 3, and 5 is caused by equilibrium between processes acting in opposite directions; settling to the bottom of suspension contaminated by radionuclide and a shift of radionuclide from solution to particles in a mineral suspension, as well as assimilation of radionuclide from solution by plankton. It is likely that over the part of the river from section 5 to section 2, settling of the contaminated suspension to the bottom owing to the slowing down of the flow predominates over the processes causing the radionuclide to shift from solution into the suspension.

Table 3. Discharge and Transport of radionuclide at Sections 1 and 2 calculated from measurements in the Summer of 1975, GBq/yr

| Radionuclide | Discharge | Section | | $T_1`/T_2`$ |
| --- | --- | --- | --- | --- |
| | | $T_1`$ | $T_2$ | |
| $^{32}P$ | 488·10$^3$ | 373·10$^3$ | - | - |
| $^{51}Cr$ | 374·10$^3$ | 349·10$^3$ | 230·10$^3$ | 0.66 |
| $^{137}Cs$ | 433 | 830 | 1170 | 1.41 |
| $^{60}Co$ | 1077 | 770 | 570 | 0.74 |
| $^{58}Co$ | 7807 | 8300 | 6000 | 0.72 |
| $^{54}Mn$ | - | 1970 | 1500 | 0.76 |
| 46Sc | - | 2600 | 1960 | 0.75 |
| 65Zn | - | 5300 | 1990 | 0.38 |

Figure 1 shows the variation in the volume activity of the water as the flow moves toward the mouth of the Yenisei observed in the summer of 1981. Since it has been found [8] that there is a close correlation between the amounts of $^{58}$Co, $^{46}$Sc, $^{54}$Mn, and $^{65}$Zn in the water, the average of these was taken. The change in the volume activity of the suspension as it moves along the channel is well approximated by the exponential form

$$Y_1 = 1.8\exp(-0.0024X),$$

where $Y_1$ is the relative volume activity of radionuclide in the river water and X is the distance (km) along the river channel from Krasnoyarsk.

The dependence of the volume activity on the distance is similar for the radionuclide which migrate predominantly in the liquid phase, $^{51}$Cr and $^{32}$P (Fig.2). For $^{51}$Cr, this dependence has the form $Y_2 = 8230\exp(-0.0021X)$, where $Y_2$ is the volume activity (Bq/m$^3$) in the water. Note that the exponents characterizing the migration of impurities in the suspension and the liquid phase differ by less than 15% for the Yenisei.

In order to evaluate the role of removing radionuclide from the flow, their transport through the transverse cross section of the river was determined ata some at fixed cross sections, or points along the river. At those points where the activity is uniformly distributed over the transverse cross section of the river, the amount of transport of any radionuclide is equal to the product of its total concentration and the flow rate of water through the given cross section. A uniform distribution is observed at sections 1 and 2, owing to the influence of the Kazachinskii and Osinovskii Rapids, respectively (see Table 3).

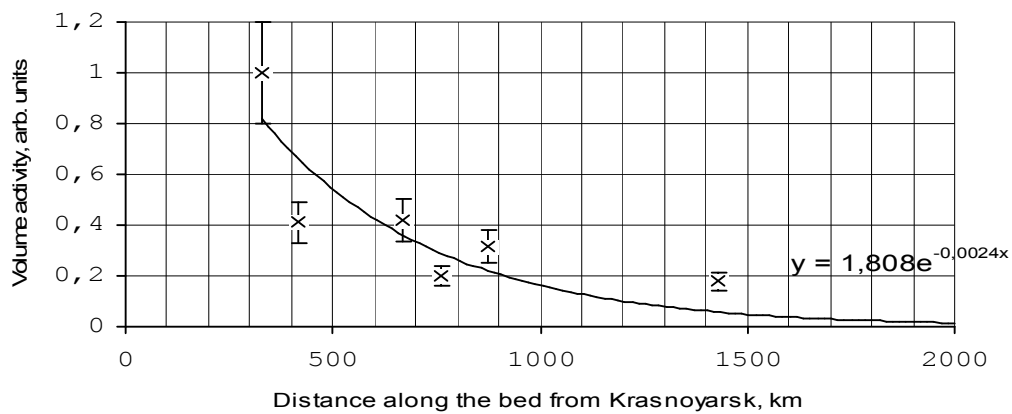

Fig.1. Amounts of $^{60}$Co, $^{46}$Sc, $^{54}$Mn, and $^{65}$Zn migrating predominantly in the solid phase in the waters of the Yenisei at different distances from Krasnoyarsk.

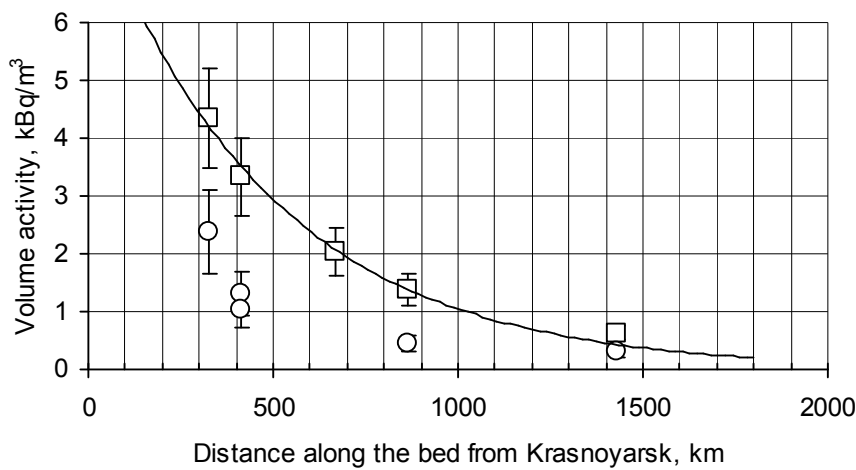

Fig.2. Amounts of $^{51}$Cr (□) and $^{32}$P (○) migrating predominantly in the liquid phase in the waters of the Yenisei in 1981.

The annual transport is estimated assuming that the rate of entry of radioactive waste into the river does not vary over a year. The water flow rates in the channel during the time of the measurements at sections 1 and 2 were 3290 and 10000m$^3$/sec, respectively. In the calculations, the decay of the radionuclide over the times they moved from the discharge point to section 1 (from velocity 5 km/h) and from section 1 to section 2 (at a velocity 4 km/h) was taken into account.

Table 3 shows that the transport of $^{51}$Cr and $^{58}$Co through section 1 is approximately equal to their discharge amounts. The transport of $^{137}$Cs through section 1 greatly exceeds its discharge. In the segment of roughly 600 km in length between section 1 and 2, during the period of the observations (August 1975), the radioactivity of the water decreased by about 25%. This was probably caused by setting of suspensions contaminated with radionuclide on the bottom. According to the observations of 1979, the radioactivity of the water in this segment of the channel decreased by about 10%, and in 1981thr transport of radionuclide through section 2 was 15% greater than through section 1; that is, during this 1981 period, radionuclide were washed out of the bottom sediments faster than they were deposited. In all the years of observations, the transport of $^{137}$Cs increased toward the mouth of the river (Table 4). The transport owing to contamination of the water by $^{137}$Cs from global fallout was calculated by multiplying the corresponding flow rate of the water by its average volume activity in the river water. This was 0.2 Bq/m$^3$, a value obtained by analyzing samples of water from the Podkamennaya Tunguska, Angara and Yenisei (near Krasnoyarsk) Rivers [9]. Table 4 implies that during the time when the direct-flow reactors were in use, the contribution from global fallout to the migration through section 2 was no more than 11%; hence, the bulk of this radionuclide transported by the Yenisei originated in erosion of the accumulation in the channel from earlier years. We neglect washout from contaminated parts of the floodplain at the time of the anomalously high water levels during the spring flood of 1966, since such high water levels were not observed in the Yenisei in subsequent years because the river's flow was regulated after construction of the Krasnoyarsk Hydroelectric Power Plant in 1967.

Table 4. Transport of $^{137}$Cs in Different Years at Various Distances from Krasnoyarsk, GBq/yr

| Population centre | Section | Distance, km | Contamination of the river water, total/caused by global fallout | | | | |
|---|---|---|---|---|---|---|---|
| | | | 1975 | 1976 | 1981 | 1985 | 2001 |
| Kazachinskoe | 1 | 266 | 830/20 | 800/20 | 470/20 | - | 120/30 |
| Shirokii Log | 3 | 327 | 910/20 | - | 470/20 | 700/30 | 140/30 |
| Bor | 2 | 864 | 1170/60 | 1100/50 | 830/50 | 690/70 | 200/60 |
| Turukhansk | | 1446 | - | - | 900/60 | 1370/80 | 120/60 |

Estimates based on taking and analyzing soil samples showed that the amount of $^{137}$Cs in a part of the floodplain near the town of Kazachinskoe in 1975 and 2000 decreased only because it underwent radioactive decay; i.e., to within the limits of error of the estimates of no washout from floodplain contamination resulting from the operation of the Complex was found [8].

The data of Table 4 were used to calculate the washout of $^{137}$Cs in different years (the contribution owing to global fallout is eliminated) from the segments of the channel between sections 1 and 2 and from section 1 to Turukhansk (Table 5).

The transport of $^{137}$Cs through sections 1 and 3 is almost identical, since the distance between these sections is only 61 km. Given that the washout is defined as the difference of two quantities with a measurement error of 20-30%, the relative computational error may be as high as 50%.

Assuming that the washout of $^{137}$Cs is directly proportional to its content of the segment of the channel under study between sections 2 and 1, we can write

$$\Delta_{21} = T_2 - T_1 = kS_{21},$$

where $T_2$ and $T_1$ are the transport of $^{137}$Cs through sections 2 and 1, respectively; $S_{21}$ is the amount of $^{137}$Cs in the segment of the channel between sections 2 and 1; and $k<1$ is a coefficient of proportionality.

The drop in the rate of discharge of $^{137}$Cs in the river channel can be described by the equation

$$dS(t)/dt = -\mu S(t) + CF, \qquad (1)$$

where S(t) is the amount in the river channel (GBq), µ is the self-cleaning constant (yr$^{-1}$); C is the discharge rate (GBq/yr); and F is the fraction that settles in the channel.

The solution of Eq.(1) for constant discharge rate and fraction of $^{137}$Cs settling in the channel is

$$S=S_0\exp(-\mu t)+CF[(1-\exp(-\mu t))]/\mu; \qquad (2)$$

where $S_0$ is the $^{137}$Cs content at t=0.

Assuming that the washout D of $^{137}$Cs is proportional to the content S, i.e., D=kS, and multiplying Eq.(2) by k, we obtain

$$D=D_0\exp(-\mu t)+CFk[(1-\exp(-\mu t))]/\mu. \qquad (3)$$

Based on Eq.(3), a system of equations

$$D(1975)=D_0\exp(-\mu)+C(1975)Fk[(1-\exp(-\mu))]/\mu;$$
$$D(1976)=D(1975)\exp(-\mu)+C(1976)Fk[(1-\exp(-\mu))]/\mu; \qquad (4)$$
$$\ldots\ldots\ldots\ldots\ldots\ldots\ldots\ldots\ldots\ldots\ldots\ldots\ldots$$
$$D(2001)=D(2000)\exp(-\mu)+C(2001)Fk[(1-\exp(-\mu t))]/\mu$$

can be set up which makes it possible to calculate the washout of $^{137}$Cs for each year from 1975 through 2001, given a different initial washout, self cleaning constant, fraction of $^{137}$Cs settling in the channel, and proportionality constant. By comparing of calculated values and the measured $\Delta_i$, an optimum initial washout, self cleaning constant, fraction of $^{137}$Cs settling in the channel, and proportionality constant can be selected for which the functional $\Sigma(D_i-\Delta_i)^2$ is minimal.

The mean annual washout of radionuclide may vary from the value calculated using the observational data for one or two summer since the flow of water is greater in the spring and lower in the winter and autumn than during the summer low water; hence the optimization of the parameters of Eq. (3) was done for some various in the observed washout. The results of the optimization are listed in Table 6, which shows that when the observed washout rate is varied by a factor of 3 (from 0.5 to 1.5), the optimized self cleaning constant µ varies by only 20% and is equal to 0.19 yr$^{-1}$ on the average. For a conserved (nondecaying) impurity, the self cleaning constant is

$$\mu_k = \mu-\lambda(^{137}Cs) = 0.19-0.0231 = 0{,}17 \text{ yr}^{-1},$$

where $\lambda(^{137}Cs)$ is the radioactive decay constant for $^{137}$Cs; for $^{60}$Co, with a half life of 5,27 years, the self cleaning constant becomes

$$\mu(^{60}Co) = \mu_k+\lambda(^{60}Co) = 0.17+0{,}132 = 0{,}302 \text{ yr}^{-1},$$

where $\lambda(^{60}Co)$ is the radioactive decay constant for $^{60}$Co.

From 1975 through 1979, the transport of $^{137}$Cs was roughly constant (see Table 5).

Table 5. Washout of $^{137}$Cs from Sections of the Yenisei Channel, GBq/yr

| Channel section | 1975 | 1979 | 1981 | 1985 | 2001 |
|---|---|---|---|---|---|
| From section 1 (3) to section 2 | 300 | 270 | 330 | −50* | 50 |
| From section 1 (3) to Turukhansk | - | - | 390 | 620 | −40* |
| *Accumulation | | | | | |

If it is assumed that the discharge rate also was constant and equal to the average for these years, then k=µ. Under these conditions, we can estimate the fraction of $^{137}$Cs that settled in the channel. These estimates show that over this segment of the channel, with a length of about 600 km, from 3 to 7% of the $^{137}$Cs could be converted into bottom sediments (see Table 6).

Table 6. Optimized Parameters

| Washout rate relative to observed | D(1975), GBq | µ, yr$^{-1}$ | Fk | F (for k= µ) |
|---|---|---|---|---|
| 0.5 | 170 | 0.21 | 0.006 | 0.027 |
| 0.7 | 230 | 0.2 | 0.008 | 0.04 |
| 1.0 | 330 | 0.18 | 0.009 | 0.052 |
| 1.5 | 500 | 0.17 | 0.011 | 0.066 |

The amount of radionuclide in the channel of the Yenisei was determined directly in 1972 and 1973 by taking samples of soft bottom sediment and analyzing them in a fixed laboratory (Table 7). Concentration equilibrium between the radionuclide contents in the water and bottom sediments collected at a single point was found to be lacking. An analysis shows that, because of the reduced discharge, self cleaning of the channel, and radioactive decay, the amount of $^{137}$Cs in the bottom sediments of the Yenisei felt by more than a factor of 10 to 70 GBq from 1972 to 2001 [8].

Table 7. Radionuclide Content in the Channel of the Yenisei According to Samples Taken in 1972, GBq

| Radionuclide | Channel segment: distance along the channel from Krasnoyarsk, km | | | |
|---|---|---|---|---|
| | 87-267 | 267-864 | 870-2000 | 87-2000 |
| $^{137}$Cs | 110 | 900 | 4300 | 5300 |
| $^{60}$Co | 100 | 740 | 990 | 1800 |
| $^{65}$Zn | 360 | 1870 | 2600 | 4800 |
| $^{152}$Eu | 55 | 400 | 670 | 1100 |
| $^{54}$Mn | 180 | 460 | 1100 | 1700 |

Let us compare the $^{137}$Cs content in the bottom sediments in the segment of the channel from 267 to 864 km (see Table 7) with estimates derived from an analysis of its transport by the water flow. Figure 3a shows the $^{137}$Cs content as a function of time calculated using Eqs. (3) and (4) and the parameters of Table 6 for observed washout rates of 1 (Δ), 0.7 (×), and 0.5 (+). Since there are no data on the discharges before 1975, it was assumed that from 1970 through 1974 the discharge rate was equal to the average over 1975-1979 (3200 GBq/yr). figure 3 shows that the data of Table 7 with a 30% relative error taken into account, are consistent with the calculated curve for a washout rate of 0.5. A similar comparison for $^{60}$Co is shown in Fig. 3b. In the calculations, it was assumed that from 1970 through 1974 the discharge rate was equal to the average over 1975-1979 (1630 GBq/yr) and that the fraction of $^{60}$Co that settles in the channel is the same as that $^{137}$Cs.

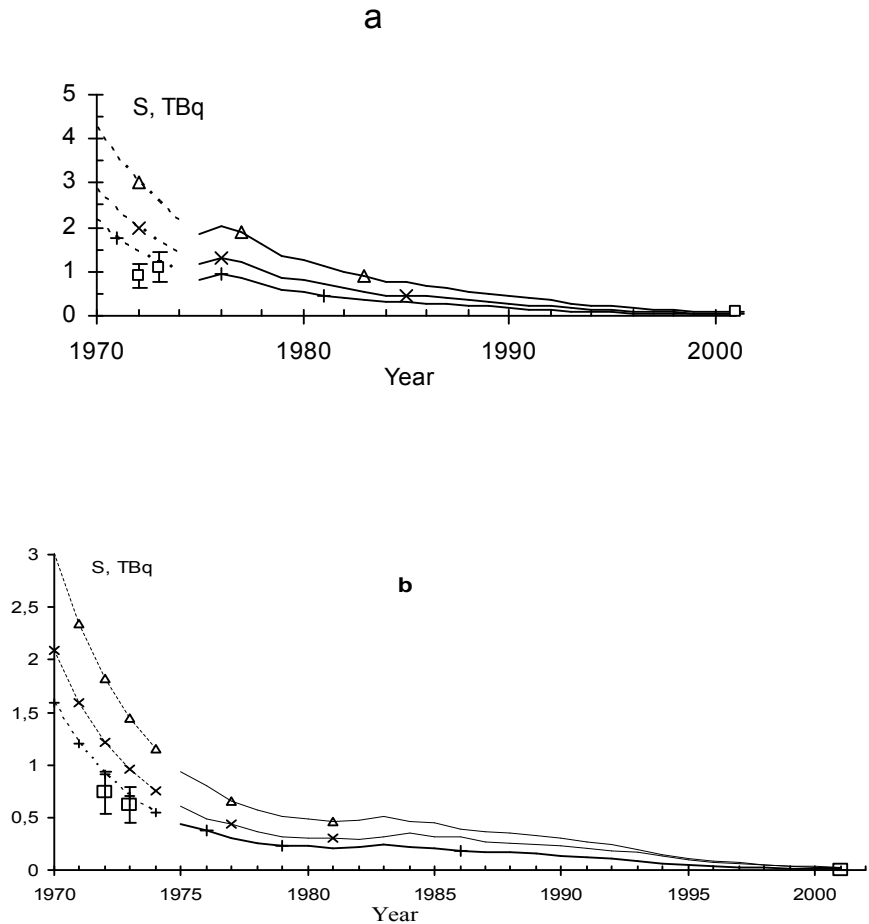

Fig.3. Comparison of the amount of $^{137}$Cs (a) and $^{60}$Co(b) in bottom sediments of the Yenisei channel between segments 1 and 2 in 1972, 1973, and 2001 with the calculated stock using the radionuclide transport (average for all seasons of the year). The calculation has carried out for rates of the washout 1(Δ); 0,7(×); and 0,5(+) relatively the observe washout (transport) in summer.

Observations of the contamination of the Yenisei thus yield data on the radionuclide content in the water and bottom sediments at different distances from the source of the effluent, the Mining –Chemical Complex. The amounts of radionuclide in the channel in 1972 and 1973, and the way these amounts varied through 2001, have been estimated. The amounts migration in the Yenisei waters have been studied, in particular, the distribution of radionuclide over the solid and liquid phases. The advantages of using the parameter f, the fraction of radionuclide in suspension, to characterize the phase state of the radionuclides in natural waters, rather than the traditionally employed suspension-water distribution coefficient, are pointed out.

By analyzing the transport of $^{137}$Cs in a segment of the channel with a length of nearly 600 km, it has been possible to evaluate the constant, characteristic tendency of the channel to clean itself (averaged over all seasons of the year) of manmade industrial impurities. For a conserved impurity, the cleaning rate is approximately 0.19 yr$^{-1}$, which corresponds to a half life of 3.6 yr for the contaminants.

In 2001, this work was supported by the International Science and Technology Center (ISTC) (Project 1404).